\documentclass[11pt]{article}

\usepackage[margin=1in]{geometry}
\usepackage[T1]{fontenc}
\usepackage[utf8]{inputenc}
\usepackage{lmodern}
\usepackage{microtype}
\usepackage{hyperref}
\usepackage{xurl}
\usepackage{booktabs}
\usepackage{longtable}
\usepackage{tabularx}
\usepackage{enumitem}
\usepackage{amsmath}
\usepackage{xcolor}
\usepackage{tikz}
\usetikzlibrary{arrows.meta,positioning,fit,backgrounds}

\hypersetup{
  colorlinks=true,
  linkcolor=blue!50!black,
  citecolor=blue!50!black,
  urlcolor=blue!50!black,
  pdftitle={OpenClaw PRISM: A Zero-Fork, Defense-in-Depth Runtime Security Layer for Tool-Augmented LLM Agents}
}

\title{OpenClaw PRISM: A Zero-Fork, Defense-in-Depth Runtime Security Layer for Tool-Augmented LLM Agents}
\author{Frank Li\\UNSW Sydney\\\texttt{research@n1a.net}}
\date{}

\begin{document}

\maketitle

\begin{abstract}
Tool-augmented LLM agents introduce security risks that extend beyond user-input filtering, including indirect prompt injection through fetched content, unsafe tool execution, credential leakage, and tampering with local control files. We present OpenClaw PRISM, a zero-fork runtime security layer for OpenClaw-based agent gateways. PRISM combines an in-process plugin with optional sidecar services and distributes enforcement across ten lifecycle hooks spanning message ingress, prompt construction, tool execution, tool-result persistence, outbound messaging, sub-agent spawning, and gateway startup. Rather than introducing a novel detection model, PRISM integrates a hybrid heuristic-plus-LLM scanning pipeline, conversation- and session-scoped risk accumulation with TTL-based decay, policy-enforced controls over tools, paths, private networks, domain tiers, and outbound secret patterns, and a tamper-evident audit and operations plane with integrity verification and hot-reloadable policy management. We outline an evaluation methodology and benchmark pipeline for measuring security effectiveness, false positives, layer contribution, runtime overhead, and operational recoverability in an agent-runtime setting, and we report current preliminary benchmark results on curated same-slice experiments and operational microbenchmarks. The system targets deployable runtime defense for real agent gateways rather than benchmark-only detection.
\end{abstract}


\section{Introduction}

Tool-augmented LLM agents expand the attack surface of language-model systems beyond what conventional input filtering can address. In a modern agent runtime, untrusted instructions may arrive not only through direct user prompts but also through retrieved web content, tool-returned text, intermediate prompt construction steps, outbound message generation, and modifications to local control files. Once an agent acquires the ability to browse the web, invoke shell commands, persist intermediate artifacts, and emit externally visible output, security risks---prompt override, tool abuse, credential leakage, control-file tampering---become distributed across the entire runtime rather than confined to a single input boundary \cite{securityofaiagents2024,asb2024,owaspllmtop102025,mitreatlas2023}.

Existing defenses, however, still operate predominantly at one or two checkpoints, typically by classifying incoming prompts or filtering final outputs. While these boundary-focused approaches remain useful, they are structurally mismatched to the execution model of tool-using agents. Indirect prompt injection, for instance, may arrive through fetched pages or tool-returned results long after the original user prompt has passed inspection \cite{greshake2023,llamafirewall2025,indirectpromptfirewalls2025}. Low-grade suspicious signals may accumulate across multiple conversational turns before crossing a meaningful risk threshold. Tool misuse and outbound credential exfiltration often demand policy enforcement rather than text classification alone \cite{safe-tool-use2026}. For agent gateways, the central challenge is therefore not merely whether a single text string appears malicious, but how security controls can be distributed coherently across a multi-stage runtime that interleaves prompts, tool invocations, mutable state, and operator workflows.

This challenge is as much an engineering problem as it is a modeling problem. In practice, runtime defenses must be deployable without invasive modifications to the host framework, maintainable as that framework evolves, and observable by the operators who depend on them. Security logic that requires forking the upstream codebase becomes a maintenance burden with every upstream release. Static policies that cannot be reloaded, inspected, or verified at runtime are difficult to operate under real conditions. Detection-only approaches are likewise insufficient for production gateways, where operators need explicit enforcement points, auditable security events, and recovery workflows for handling policy updates and exceptions. This emphasis on deployment-time governance is also consistent with broader AI risk-management guidance that treats safety as a socio-technical operational problem rather than a model-only property \cite{nistairmf2023}. A practical agent defense system must therefore function as an operational runtime layer---not as a benchmark-only detector.

We present OpenClaw PRISM, a zero-fork runtime security layer for OpenClaw-based tool-augmented LLM agent gateways. PRISM combines an in-process plugin with optional sidecar services and distributes enforcement across ten lifecycle hooks covering message ingress, prompt construction, tool execution, tool-result persistence, outbound messaging, sub-agent spawning, and gateway startup. Rather than introducing a novel detection model, PRISM integrates several complementary defense mechanisms within a unified runtime architecture:

\begin{itemize}[leftmargin=1.5em]
  \item \textbf{Lifecycle-wide interception} across ten hooks that span the full agent interaction cycle.
  \item \textbf{A hybrid scanning pipeline} that applies fast heuristic scoring first and can escalate suspicious tool results to an LLM-assisted classifier.
  \item \textbf{Session-level and conversation-level risk accumulation} with time-decaying state and graduated response thresholds.
  \item \textbf{Policy-enforced controls} over tool invocations, protected paths, private networks, domain tiers, and outbound secret patterns.
  \item \textbf{A tamper-evident audit and operations plane} with chained audit records, integrity verification, hot-reloadable policy configuration, and dashboard allow workflows.
\end{itemize}

The goal is deployable runtime defense: a system that can be attached to an existing gateway, supervised as long-running services, audited after the fact, and tuned by operators---all without forking the upstream framework.

This paper makes the following contributions:

\begin{enumerate}[leftmargin=1.5em]
  \item \textbf{Zero-fork runtime defense architecture.} We design and implement a runtime defense layer for OpenClaw-based agent gateways that combines an in-process plugin with optional sidecar services, avoiding any modification to upstream gateway code.
  \item \textbf{Lifecycle-wide enforcement with staged risk response.} We implement enforcement across ten runtime hooks together with a hybrid heuristic-first, LLM-assisted scanning path and thresholded risk accumulation that supports graduated responses rather than one-shot classification.
  \item \textbf{Unified tool and network governance.} We unify tool governance, protected-path enforcement, private-network and domain controls, and outbound secret filtering within a single runtime security layer for tool-using agents.
  \item \textbf{Tamper-evident audit and operational plane.} We provide an operational security plane that includes chained audit records with integrity verification, component health probing, hot-reloadable policy configuration, and dashboard allow workflows.
  \item \textbf{Evaluation methodology and benchmark artifacts.} We outline an evaluation methodology and accompanying benchmark artifact structure for measuring security effectiveness, false positives, layer contribution, runtime overhead, and operational recoverability in an agent-runtime setting.
\end{enumerate}

The remainder of this paper is organized as follows. Section~2 defines the agent runtime model and threat model. Section~3 presents the PRISM architecture and its core defense mechanisms. Section~4 describes the implementation and deployment model. Section~5 outlines the evaluation methodology and reports current preliminary benchmark results. Section~6 positions PRISM relative to related work, and Section~7 discusses limitations and future directions.

\section{Background and Threat Model}

\subsection{Agent Runtime Model}

PRISM is designed for a tool-augmented gateway runtime in which an LLM-driven agent interacts with users, constructs prompts, invokes tools, processes tool-returned content, and emits externally visible responses. In this setting, security-relevant information does not flow through a single input channel. Instead, it traverses several stages that are all potentially attackable: user messages, retrieved external content, intermediate prompt construction, tool invocation parameters, tool-returned artifacts, outbound messages, and selected local control files.

The runtime entities relevant to PRISM are the user, the OpenClaw gateway, the in-process PRISM plugin, and a set of optional sidecar services. The sidecars include a scanner service for remote classification of suspicious tool-returned content, an invoke-guard proxy for policy-backed tool governance, a dashboard for audit inspection and policy management, and a file monitor for integrity checks over selected local files. The plugin serves as the point of lifecycle interception inside the gateway; the sidecars extend scanning, enforcement, operations, and monitoring capabilities without requiring a fork of the upstream framework.

At a high level, the runtime proceeds as follows. A user message enters the gateway and may be inspected before prompt construction. The gateway then prepares the model-facing prompt, during which security context may be injected if earlier stages have accumulated sufficient risk. The agent may subsequently invoke tools, whose parameters and destinations can be checked before execution. Tool results may then be scanned, sanitized, persisted, or incorporated into later model context. When the gateway prepares an outbound response, the final content can be checked again for residual injection signals, secret leakage, or elevated conversation risk before it is sent externally. This staged execution model is the background assumption that motivates PRISM's lifecycle-wide defense design.

\subsection{Protected Assets}

We identify five classes of assets as security-relevant within this runtime model. The first is hidden control context, including system prompts and other internal instructions that shape agent behavior but are not intended to be exposed or overridden by untrusted content. The second is tool execution privilege: the ability to invoke tools, reach selected network destinations, or perform actions whose effect extends beyond text generation. The third is local configuration and control state, including security-relevant files whose modification could weaken the gateway's policy boundary. The fourth is secret material, such as API tokens, credentials, and other sensitive values that could be disclosed through tool results or outbound messages. The fifth is security telemetry itself, including audit records and mutable policy state that operators rely on to inspect and adjust the system.

These assets matter because attacks against tool-using agents are rarely limited to prompt misclassification. A successful adversary may instead attempt to induce unsafe tool execution, cause the agent to disclose sensitive material, poison persistent state, or tamper with the control and audit surfaces that operators depend on during incident response and recovery.

\subsection{Adversaries and Goals}

PRISM considers adversaries with four principal capabilities. First, an attacker may directly control user-provided text and attempt to override instructions, exfiltrate hidden context, or induce unsafe tool use through ordinary conversational input. Second, an attacker may control external content consumed by the agent---such as fetched web pages, API responses, or other tool-returned text---and use that content to deliver indirect prompt injection after the original user message has already passed inspection. Third, an attacker may attempt to steer outbound behavior by causing the agent to transmit sensitive content to risky or untrusted destinations. Fourth, a local attacker with file-level access to selected paths may attempt to modify protected control files or interfere with the evidence trail used for auditing and incident response.

Within this model, the adversary goals that PRISM addresses are prompt override, policy bypass, unsafe tool execution, secret exfiltration, long-horizon escalation through repeated low-grade signals, and control-file tampering. These attack classes overlap with practitioner-oriented taxonomies such as the OWASP Top 10 for LLM Applications and MITRE ATLAS, but PRISM narrows them to the subset that a gateway-attached runtime defense can realistically mediate \cite{owaspllmtop102025,mitreatlas2023}. The long-horizon case is particularly important because some attacks do not appear decisively malicious at any single checkpoint. Instead, they emerge as a pattern of suspicious signals distributed across multiple turns, tools, or external interactions. This observation is one reason the paper emphasizes risk accumulation and thresholded runtime response rather than single-shot classification alone.

\subsection{Out-of-Scope Attacks and Assumptions}

PRISM does not attempt to address every security problem relevant to LLM agents. We treat model poisoning, training-time corruption, and arbitrary compromise of model internals as out of scope. We also exclude full host compromise, kernel- or hardware-level attacks, and arbitrary supply-chain compromise outside the PRISM deployment boundary. PRISM is not intended to serve as a complete OS sandbox or a general-purpose outbound firewall for all processes on the host.

The paper therefore assumes that the gateway process, the host environment, and the deployment stack retain a baseline level of integrity beyond what PRISM itself enforces. Under that assumption, PRISM aims to improve runtime security at the agent-gateway layer by making attacks more detectable, more interruptible, and more auditable across the stages it directly mediates.

\section{System Design}

\subsection{Architecture Overview}

PRISM adopts a zero-fork integration strategy for OpenClaw-based agent gateways. The core principle is to insert security logic at the gateway lifecycle layer without modifying upstream gateway source code. To achieve this, PRISM combines an in-process plugin with a small set of optional sidecar services. The plugin is responsible for runtime interception inside the gateway itself, while the sidecars extend the system with external scanning, tool-governance mediation, audit inspection, policy management, and file monitoring. This separation allows PRISM to remain closely tied to actual gateway execution while keeping heavier operational functionality out of the plugin's critical path.

The plugin is the architectural center of the system. It registers lifecycle hooks that observe or intercept message ingress, prompt construction, tool execution, post-tool handling, outbound messaging, sub-agent creation, session cleanup, and gateway startup. Around this hook surface, PRISM can accumulate risk signals, block unsafe actions, sanitize suspicious content, and emit audit events. The sidecars provide specialized functions that are not naturally embedded inside the gateway process: the scanner service performs authenticated remote classification of suspicious tool-returned content, the invoke-guard proxy provides policy-backed control over selected tool invocations, the dashboard exposes audit and configuration workflows, and the monitor watches selected files for integrity-relevant changes.

This architecture deliberately separates data-plane enforcement from operational support surfaces. The plugin and proxy participate directly in security decisions that affect live execution. The scanner contributes additional classification signal, but it is not the sole enforcement mechanism and is not treated as a universally trusted oracle. The dashboard and monitor support operator visibility, configuration management, and recovery. Audit storage forms the evidence plane that ties these components together through append-only records and later verification.

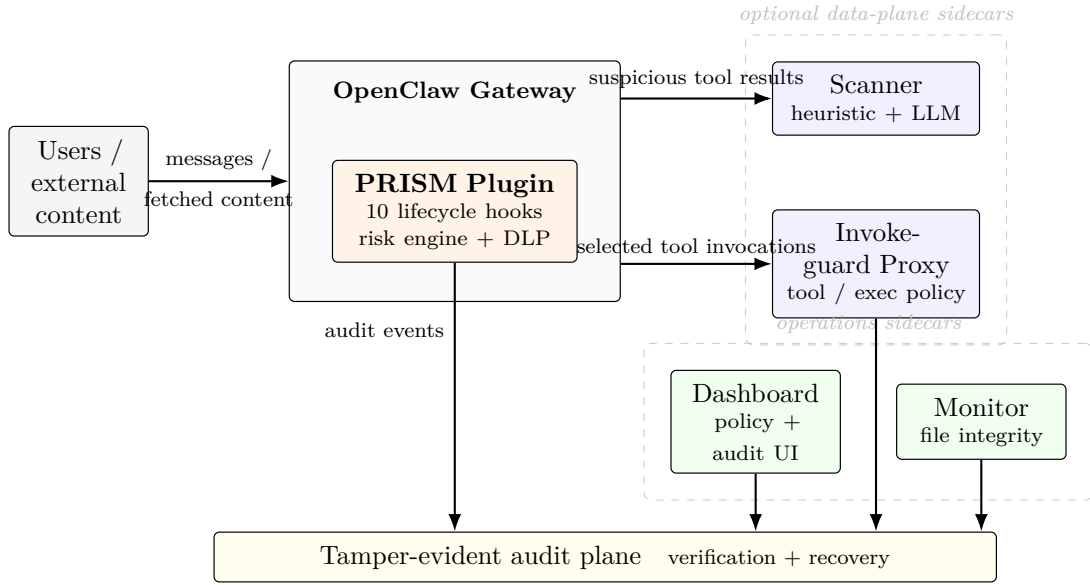
\begin{figure}[t]
\centering
\begin{tikzpicture}[
  font=\small,
  >=Latex,
  box/.style={draw, rounded corners=2pt, align=center, inner sep=5pt, line width=0.45pt},
  flow/.style={->, thick},
  lbl/.style={font=\scriptsize}
]

\node[box, fill=gray!8, text width=1.5cm] (user) at (-0.8, 2.5)
  {Users /\\external\\content};

\node[box, fill=gray!4, minimum width=4.4cm, minimum height=3.2cm] (gw) at (4.2, 2.5) {};
\node[font=\footnotesize\bfseries, anchor=north] at ([yshift=-4pt]gw.north)
  {OpenClaw Gateway};

\node[box, fill=orange!10, text width=2.9cm] (plugin) at (4.2, 2.1)
  {\textbf{PRISM Plugin}\\[-2pt]
   {\scriptsize 10 lifecycle hooks}\\[-2pt]
   {\scriptsize risk engine + DLP}};

\node[box, fill=blue!6, text width=2.4cm] (scanner) at (9.8, 3.6)
  {Scanner\\[-2pt]{\scriptsize heuristic + LLM}};
\node[box, fill=blue!6, text width=2.4cm] (proxy) at (9.8, 1.4)
  {Invoke-guard Proxy\\[-2pt]{\scriptsize tool / exec policy}};

\begin{scope}[on background layer]
  \node[draw, dashed, rounded corners=3pt, inner sep=10pt,
    draw=gray!40, line width=0.4pt, fit=(scanner)(proxy)] (dpgrp) {};
\end{scope}
\node[font=\scriptsize\itshape, text=gray!50, anchor=south] at (dpgrp.north)
  {optional data-plane sidecars};

\node[box, fill=green!6, text width=1.9cm] (dashboard) at (8.2, -0.7)
  {Dashboard\\[-2pt]{\scriptsize policy + audit UI}};
\node[box, fill=green!6, text width=1.9cm] (monitor) at (11.2, -0.7)
  {Monitor\\[-2pt]{\scriptsize file integrity}};

\begin{scope}[on background layer]
  \node[draw, dashed, rounded corners=3pt, inner sep=10pt,
    draw=gray!40, line width=0.4pt, fit=(dashboard)(monitor)] (opsgrp) {};
\end{scope}
\node[font=\scriptsize\itshape, text=gray!50, anchor=south] at (opsgrp.north)
  {operations sidecars};

\node[box, fill=yellow!8, minimum width=10.4cm, minimum height=0.65cm] (audit) at (6.2, -2.5)
  {Tamper-evident audit plane\quad{\scriptsize verification + recovery}};


\draw[flow] (user.east) --
  node[above, lbl] {messages /}
  node[below, lbl] {fetched content}
  (gw.west);

\draw[flow] (gw.east |- scanner) --
  node[above, lbl, midway] {suspicious tool results}
  (scanner.west);

\draw[flow] (gw.east |- proxy) --
  node[above, lbl, midway] {selected tool invocations}
  (proxy.west);

\draw[flow] (plugin.south) --
  node[left, lbl, pos=0.25] {audit events}
  (plugin.south |- audit.north);

\draw[flow] (proxy.south) -- (proxy.south |- audit.north);

\draw[flow] (dashboard.south) -- (dashboard.south |- audit.north);

\draw[flow] (monitor.south) -- (monitor.south |- audit.north);

\end{tikzpicture}
\caption{PRISM architecture. The in-process plugin enforces security decisions inside the OpenClaw gateway. Optional data-plane sidecars extend scanning and tool governance; operations sidecars provide policy management and file-integrity monitoring. The audit plane ties runtime decisions to later verification and recovery.}
\label{fig:architecture}
\end{figure}

\subsection{Lifecycle-Wide Enforcement}

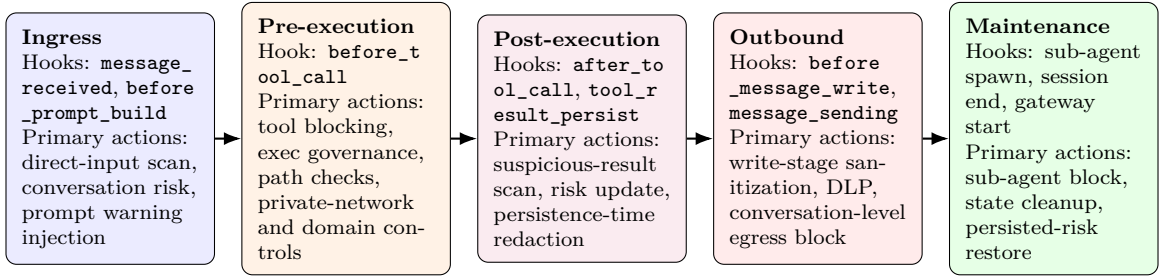
\begin{figure}[t]
\centering
\begin{tikzpicture}[
  font=\scriptsize,
  >=Latex,
  phase/.style={draw, rounded corners, align=left, text width=2.35cm, minimum height=2.55cm, inner sep=6pt},
  arrow/.style={->, thick}
]
\node[phase, fill=blue!8] (ingress) {\textbf{Ingress}\\Hooks: \path{message_received}, \path{before_prompt_build}\\Primary actions: direct-input scan, conversation risk, prompt warning injection};
\node[phase, fill=orange!10, right=0.35cm of ingress] (pre) {\textbf{Pre-execution}\\Hook: \path{before_tool_call}\\Primary actions: tool blocking, exec governance, path checks, private-network and domain controls};
\node[phase, fill=purple!8, right=0.35cm of pre] (post) {\textbf{Post-execution}\\Hooks: \path{after_tool_call}, \path{tool_result_persist}\\Primary actions: suspicious-result scan, risk update, persistence-time redaction};
\node[phase, fill=red!8, right=0.35cm of post] (outbound) {\textbf{Outbound}\\Hooks: \path{before_message_write}, \path{message_sending}\\Primary actions: write-stage sanitization, DLP, conversation-level egress block};
\node[phase, fill=green!10, right=0.35cm of outbound] (maint) {\textbf{Maintenance}\\Hooks: sub-agent spawn, session end, gateway start\\Primary actions: sub-agent block, state cleanup, persisted-risk restore};

\draw[arrow] (ingress.east) -- (pre.west);
\draw[arrow] (pre.east) -- (post.west);
\draw[arrow] (post.east) -- (outbound.west);
\draw[arrow] (outbound.east) -- (maint.west);
\end{tikzpicture}
\caption{Lifecycle-wide enforcement in PRISM. Security controls are distributed across five runtime phases, allowing the system to escalate from observation and warning to hard policy enforcement, redaction, and stateful recovery.}
\label{fig:lifecycle}
\end{figure}

The central systems claim of PRISM is that agent security should be distributed across the runtime lifecycle rather than concentrated at a single boundary. PRISM operationalizes this principle through ten hooks grouped into five phases, summarized in Figure~\ref{fig:lifecycle}.

The \textbf{ingress phase} consists of \texttt{message\_received} and \texttt{before\_prompt\_build}. These hooks allow the system to inspect direct user-visible content before it becomes part of later prompt state. The first hook is conversation-oriented: when inbound text appears suspicious, PRISM can attach risk to a conversation-scoped key. The second hook is session-oriented: prompt text is scanned again, and if enough risk has accumulated, PRISM prepends a security notice warning the model not to obey instructions embedded in fetched content or tool results. This phase therefore serves two complementary roles: early detection and context-level warning before tools are involved.

The \textbf{pre-execution phase} is centered on \texttt{before\_tool\_call}, which is the principal synchronous interception point for active policy enforcement. At this stage, PRISM can block high-risk tools for elevated-risk sessions, reject shell metacharacters and trampoline forms such as \texttt{bash -c} or \texttt{python -c}, enforce executable allowlists and deny patterns, check selected file paths, reject private-network destinations, and apply domain-tier handling. This is where the system transitions most clearly from detection into explicit runtime control.

The \textbf{post-execution phase} includes \texttt{after\_tool\_call} and \texttt{tool\_result\_persist}. The first hook targets suspicious tool-returned content. For configured scan tools, PRISM first applies local heuristics to the returned text; only if that text appears suspicious does it invoke the remote scanner service. Scanner verdicts can then add session-level risk, while scanner failure is handled by adding a bounded risk signal rather than triggering a hard denial at that stage. The second hook, \texttt{tool\_result\_persist}, is synchronous and sanitizes suspicious tool output before it is written into persistent or reusable runtime state. Together, these hooks are designed to reduce the likelihood that malicious tool-returned text survives into later prompt context unchanged.

The \textbf{outbound phase} is handled by \texttt{before\_message\_write} and \texttt{message\_sending}. The former acts as a final write-stage defense against suspicious text that would otherwise be persisted or surfaced in a later response. The latter adds two additional controls: outbound DLP checks for implemented secret patterns and conversation-level risk checks that can block message emission when the surrounding conversation has accumulated excessive risk. The purpose of this phase is to ensure that even if earlier stages missed or only partially mitigated a problem, the final hop retains the ability to interrupt exfiltration or leakage.

The final group is \textbf{lifecycle maintenance}, implemented by hooks for sub-agent spawning, session end, and gateway startup. These hooks ensure that the defense system is not limited to a single interaction. PRISM can block sub-agent creation for sessions that have accumulated high risk, clear session-scoped state at session termination, and restore persisted risk state when the gateway starts. As a result, the system's policy logic can span multiple turns and survive restart when configured to do so.

\subsection{Two-Tier Injection Scanning}

PRISM's scanning design is intentionally hybrid. It does not rely on a single classifier, and it does not assume that every runtime stage requires the same depth of analysis. Instead, the system uses fast heuristic scoring broadly and reserves remote model-assisted classification for narrower cases where suspicious tool-returned content merits additional scrutiny.

The first tier is a canonicalization and heuristic pipeline implemented in shared runtime logic. It normalizes Unicode using NFKC, performs limited percent-decoding, strips zero-width characters, and collapses formatting artifacts before applying weighted detection rules. These rules capture several common injection and abuse patterns, including instruction override phrases, system-prompt exfiltration attempts, credential-exfiltration language, tool-abuse command patterns, role overrides, format-token injection, and selected obfuscation signals. PRISM supplements these pattern rules with feature-style scoring and bonuses tied to suspicious transformed content. The result is a low-latency suspiciousness score that can be reused at several hooks throughout the lifecycle.

The second tier resides in the scanner sidecar. When the plugin encounters suspicious tool-returned content from configured scan tools, it can submit that text to the scanner over an authenticated HTTP endpoint. Inside the scanner, heuristics are applied again; high-scoring results can be short-circuited as malicious without invoking the LLM judge. Otherwise, the scanner may query a local Ollama-served model with a bounded timeout and combine the model output with heuristic evidence. The final scanner verdict is mapped to \texttt{benign}, \texttt{suspicious}, or \texttt{malicious}, with graceful fallback to the heuristic-only result if the model stage fails. This design keeps the classifier role bounded: the LLM judge is a secondary signal source, not the sole security anchor of the system.

\subsection{Session Risk Engine and Policy Response}

PRISM translates local detection events into runtime policy through a risk engine with explicit scope and decay semantics. A key design choice is that PRISM does not bind all risk to a shared channel identifier. Instead, it distinguishes between conversation-scoped and session-scoped state. Inbound and outbound conversation-level logic uses a conversation identifier when one is available, whereas prompt- and tool-related signals are tied to session keys. This separation prevents unrelated sessions from sharing risk simply because they happen to traverse a common channel-like identifier.

Risk entries carry a time-to-live and are periodically swept, which allows the system to model elevated risk as a time-bounded operational condition rather than as a permanent label. When configured, risk state may also be persisted and restored across gateway restarts. The practical effect is that PRISM can respond to patterns that emerge over multiple interactions while still allowing stale signals to decay naturally.

The response policy is thresholded rather than binary. At lower risk levels, PRISM injects warning context before prompt construction to reduce the likelihood that the model follows suspicious instructions. At higher risk, PRISM blocks high-risk tools during \texttt{before\_tool\_call}. At an even higher threshold, it blocks sub-agent spawning. This staged design reflects the observation that not every suspicious signal justifies immediate hard denial, but repeated or compounding signals should eventually narrow the agent's available action surface.

\subsection{Tool, Network, and Audit Governance}

Beyond injection handling, PRISM enforces explicit governance over tool execution, selected file operations, outbound destinations, and audit integrity.

\paragraph{Tool governance.} The primary enforcement path is \texttt{before\_tool\_call}, supplemented by the invoke-guard proxy in deployments that choose to route tools through a dedicated policy service. PRISM checks executable prefixes, known-dangerous patterns, shell metacharacters, and trampoline forms that attempt to smuggle arbitrary shell execution through superficially benign tools. It also includes specific handling for Git SSH override patterns that can bypass ordinary command expectations. The goal is not to identify all unsafe commands from semantics alone, but to enforce a practical runtime policy over the tool surface that the gateway exposes.

\paragraph{Path governance.} PRISM applies canonicalized path checks to selected protected locations and can support explicit exceptions through operator-managed policy updates. This mechanism is intentionally string-level: it resolves and normalizes paths but does not claim symlink-aware filesystem containment. The design target is practical protection for known sensitive paths rather than a general replacement for filesystem sandboxing.

\paragraph{Network and outbound governance.} PRISM combines private-network blocking, tiered domain handling, and DLP checks for selected credential patterns. These controls are distributed across the runtime: risky destinations can be blocked or escalated before tool execution, while secrets can be filtered again on outbound message emission. This layering matters because agent misuse often involves both a destination choice and a payload choice; PRISM therefore treats exfiltration defense as more than a single string-matching problem.

\paragraph{Tamper-evident audit and operations.} PRISM couples runtime enforcement to a tamper-evident audit and operations plane. Audit entries include chained hash linkage and HMAC-based integrity protection, while periodic anchors support later verification. The audit path is designed to remain operationally useful even if some supporting sinks fail temporarily: enforcement does not depend on perfect audit delivery. Around the audit layer, the dashboard exposes configuration inspection, revision-aware updates, and allow workflows so that policy exceptions can be reviewed and applied through an explicit operator path rather than by ad hoc file edits. This combination of runtime control and operational feedback is central to why PRISM is best understood as a deployable security system rather than as a standalone detector.

\section{Implementation}

PRISM is implemented as a TypeScript monorepo targeting Node.js $\geq$22 and managed as a pnpm workspace. The implementation mirrors the architectural separation described in Section~3: a single in-process plugin performs lifecycle interception inside the OpenClaw gateway, while four optional sidecar services provide injection scanning, tool-invocation policy enforcement, an operations dashboard, and file-integrity monitoring. Shared utilities---including the weighted heuristics, canonicalization pipeline, audit-chain helpers, path normalization, and common type definitions---are factored into a dedicated library package, ensuring that the same security logic is used consistently across all runtime components and the evaluation harness.

The repository contains seven packages under \texttt{packages/}, each built with tsup into ESM artifacts. Table~\ref{tab:components} summarizes the major components. Line counts are approximate production TypeScript source lines under \texttt{packages/*/src}, excluding test files.

\begin{table}[ht]
\centering
\small
\begin{tabularx}{\textwidth}{l r l X}
\toprule
Package & Approx.\ LOC & Default interface & Main responsibility \\
\midrule
\texttt{dashboard} & $\sim$1,900 & HTTP \texttt{127.0.0.1:18768} & Security dashboard, audit browsing, configuration editing, allow preview/apply, component health probes \\
\texttt{plugin} & $\sim$980 & In-process OpenClaw plugin & Ten lifecycle hooks, risk accumulation, outbound controls, internal audit delegation \\
\texttt{proxy} & $\sim$610 & HTTP \texttt{127.0.0.1:18767} & Tool-invocation RBAC, session-ownership checks, dangerous-exec blocking, upstream forwarding \\
\texttt{shared} & $\sim$470 & Library (no network) & Heuristics, audit-chain primitives, path helpers, shared types \\
\texttt{cli} & $\sim$420 & Terminal entrypoint & Service start/status, verification, policy simulation, audit tail/verify \\
\texttt{scanner} & $\sim$130 & HTTP \texttt{127.0.0.1:18766} & Heuristic-first and Ollama-assisted injection scanning \\
\texttt{monitor} & $\sim$100 & File watcher (no network) & Integrity monitoring for selected local control files \\
\bottomrule
\end{tabularx}
\caption{Implementation components.}
\label{tab:components}
\end{table}

\subsection{Codebase Structure and Packaging}

The workspace is split along operational boundaries rather than purely logical module groupings. The plugin package is the only component that runs inside the gateway process; the scanner, proxy, dashboard, and monitor are standalone executables that can be supervised, upgraded, and restarted independently. This division makes deployment incremental: operators can run the plugin alone for lightweight protection, then attach sidecar services progressively as their assurance requirements grow.

The plugin exposes a declarative configuration schema (\texttt{openclaw.plugin.json}) that allows operators to tune risk TTL, protected paths, exec allow/block lists, scanner timeouts, outbound secret patterns, and domain tiers without modifying source code. PRISM is therefore not a fixed benchmark binary but a configurable runtime layer intended for integration into production gateway deployments.

Common security logic is centralized in the shared package. The weighted heuristics, canonicalization pipeline, audit-chain verification routines, and path normalization helpers are imported by the plugin, scanner, CLI, and evaluation harness alike. This centralization eliminates duplication across components and ensures that metric-oriented experiments exercise the identical code paths that would run in a live deployment.

\subsection{Deployment Model}

The reference deployment flow is driven by an installer that builds the workspace, provisions runtime secrets and policy files, and registers optional sidecar services with platform service managers. This deployment model is designed for long-running agent gateways rather than transient shell sessions. On Linux, the sidecars can be supervised through systemd; on macOS, the repository supports launchd-based supervision for the major auxiliary services together with a manual-start path where needed.

All network-exposed components default to loopback bindings. The scanner, proxy, and dashboard therefore operate as local sidecars rather than as internet-facing services. The plugin can additionally expose an optional internal audit endpoint for single-writer delegation from the dashboard; this interface is an internal loopback path, not a main plugin service port.

Importantly, the implementation does not require every component in every deployment. The plugin can operate without the dashboard or monitor, and individual sidecars are gated by environment variables (\texttt{PRISM\_SCANNER\_START}, \texttt{PRISM\_PROXY\_START}, \texttt{PRISM\_MONITOR\_START}, \texttt{PRISM\_DASHBOARD\_START}). PRISM is a modular runtime layer, not an all-or-nothing appliance.

\subsection{Operational Interfaces}

PRISM provides both machine-facing and operator-facing interfaces. Machine-facing interfaces are simple authenticated HTTP surfaces for scanning, proxy-mediated tool control, dashboard management, and component health inspection. On the operator-facing side, the CLI supports service control, status inspection, post-upgrade verification, offline policy simulation, and audit inspection and verification. These policy-simulation and audit-verification workflows are particularly noteworthy because they transform otherwise implicit runtime behavior into explicit, reproducible operator procedures, reinforcing that PRISM is designed for day-to-day operational use rather than one-time benchmark execution.

\subsection{Testing and Engineering Maturity}

The repository builds from the root with \texttt{pnpm -r build} and uses Vitest for unit and component testing. A static count over repository-owned test files yields 13 test files containing 142 test cases. These tests cover the heuristic scoring engine, audit-chain verification, plugin hook registration and lifecycle behavior, proxy policy logic, scanner classification paths, dashboard allow-action workflows, and monitor reconciliation. The test layout mirrors the component boundaries of the system: dedicated test files exist for audit-chain integrity, dashboard allow-actions, policy explanation paths, and hook-level plugin behavior. This coverage provides evidence that PRISM is implemented as a system with nontrivial operational logic rather than as a thin research prototype.

We emphasize that this test suite represents broad engineering coverage over security-critical paths, not a formal proof of security. The presence of build infrastructure, service-supervision artifacts, CLI verification commands, and component-level tests demonstrates implementation maturity, but it does not imply complete cross-platform portability, formal assurance, or production-scale validation. Those boundaries are discussed in Section~7.

\section{Evaluation}

We evaluate PRISM as a deployable runtime defense system rather than as a standalone prompt-injection classifier. This distinction is important: because PRISM distributes its security mechanisms across ten lifecycle hooks, four sidecar services, and a session-level risk engine, an evaluation limited to a single detection accuracy number would fail to capture the system's behavior. Accordingly, our evaluation is organized around five research questions:

\begin{itemize}[leftmargin=1.5em]
  \item \textbf{(RQ1) Security effectiveness.} How effectively does PRISM block representative attacks across multiple agent runtime surfaces, including prompt injection, tool abuse, credential exfiltration, and file tampering?
  \item \textbf{(RQ2) False positives.} How often does PRISM incorrectly flag benign agent workflows?
  \item \textbf{(RQ3) Layer contribution.} Which protection layers contribute meaningfully to overall security outcomes, and does removing individual layers degrade defense in measurable ways?
  \item \textbf{(RQ4) Runtime overhead.} What latency, CPU, and memory costs does PRISM introduce relative to an unprotected baseline?
  \item \textbf{(RQ5) Operational recoverability.} Does the system support practical operational recovery through audit verification, policy hot-reload, and allow workflows?
\end{itemize}

At the current stage, the benchmark infrastructure supports preliminary benchmark experiments for text classification, proxy-policy enforcement, and hook-level lifecycle behavior, while larger-scale end-to-end gateway experiments and deployment-level systems-overhead instrumentation remain future work. We therefore distinguish between (i) the target evaluation methodology that will drive the final paper results and (ii) a smaller benchmark layer used to validate the benchmark harness itself.

\begin{table}[t]
\centering
\footnotesize
\begin{tabularx}{\textwidth}{l X X X}
\toprule
RQ & Core question & Primary artifact / configuration & Principal outputs \\
\midrule
RQ1 & Security effectiveness across attack surfaces & Baseline ladder from no PRISM to full PRISM & Attack block rate, precision, recall, F1 \\
RQ2 & False positives on benign workflows & Benign suites and keyword-stress controls & False positive rate, error analysis by suite \\
RQ3 & Layer contribution and ablation impact & Single-layer removals and adjacent baseline comparisons & Delta in block rate and false positives per layer \\
RQ4 & Runtime overhead & Instrumented protected vs.\ unprotected runs & p50/p95/p99 latency, CPU overhead, memory overhead \\
RQ5 & Operational recoverability & Audit verification, policy reload, and allow-action workflows & Audit verification success, reload latency, recovery trace quality \\
\bottomrule
\end{tabularx}
\caption{Mapping from research questions to the artifacts and outputs required to answer them. This table fixes the evaluation contract in advance so that later prose remains aligned with the actual experiment design.}
\label{tab:rq-mapping}
\end{table}

\subsection{Experimental Setup}

PRISM is implemented as a TypeScript monorepo targeting Node.js and is deployed as a combination of an in-process OpenClaw plugin and four optional sidecar services: the scanner, proxy, dashboard, and file monitor. Our evaluation framework mirrors this architecture and measures security behavior at the layer where each defense mechanism is implemented, rather than treating the system as a monolithic black box.

The current benchmark harness is implemented in \texttt{run-benchmark.ts}. It recursively loads JSON-formatted test cases from two directories---\texttt{attack-corpus/} and \texttt{benign-corpus/}---executes each case against one or more evaluation engines, and emits a structured JSON report. The current harness supports eight engines:

\begin{itemize}[leftmargin=1.5em]
  \item \textbf{No-PRISM engine}, which models an unprotected gateway by forwarding requests without scanner, hook, or policy intervention.
  \item \textbf{Heuristics-only engine}, which applies the shared heuristic rules to an explicit per-case textual probe while leaving all other structure untreated.
  \item \textbf{Heuristic engine}, which exercises the shared text-canonicalization pipeline and weighted heuristic scoring rules.
  \item \textbf{Scanner engine}, which exercises the scanner's classification path, including the heuristic front-end and, when configured, the LLM cascade.
  \item \textbf{Proxy-policy engine}, which exercises token-based authentication, session-ownership checks, tool allow/deny rules, and the built-in dangerous-exec guards.
  \item \textbf{Plugin-only engine}, which executes step-sequenced lifecycle scenarios against the real plugin hooks with no functioning remote scanner.
  \item \textbf{Plugin-scanner engine}, which executes the same lifecycle scenarios while exposing the plugin's \texttt{after\_tool\_call} hook to the scanner service.
  \item \textbf{Full-PRISM engine}, which combines the plugin-scanner lifecycle path with proxy-policy enforcement on policy-layer cases.
\end{itemize}

The current preliminary corpus contains 110 hand-authored cases distributed as follows: 3 direct-injection cases, 3 indirect-injection cases, 2 exfiltration cases, 18 tool-abuse policy cases, 47 plugin-flow lifecycle cases, 15 scanner-focused contextual cases, 3 benign chat cases, 2 benign web-content cases, 15 benign policy-allowed tool-use cases, and 2 challenging benign controls designed to stress false-positive behavior. Table~\ref{tab:seed-corpus} summarizes these suites and the primary mechanism each suite is intended to exercise. This corpus remains modest relative to the full evaluation target; its immediate purpose is to validate the report structure, exercise the metrics pipeline, and isolate scanner-path and lifecycle-benchmark behavior before scaling to a larger end-to-end corpus.

\begin{table}[t]
\centering
\small
\begin{tabularx}{\textwidth}{l c c X}
\toprule
Suite family & Cases & Type & Primary mechanism exercised \\
\midrule
Direct injection & 3 & Attack & Heuristic detection of override / disclosure language \\
Indirect injection & 3 & Attack & Canonicalization, suspicious-result scanning, persistence redaction \\
Exfiltration & 2 & Attack & Secret-pattern detection and outbound safeguards \\
Tool abuse & 18 & Attack & Proxy policy checks and dangerous-exec blocking \\
Plugin-flow lifecycle cases & 47 & Mixed & Step-sequenced hook-level evaluation of plugin-only, plugin-scanner, and full-PRISM configurations \\
Scanner-focused contextual cases & 15 & Mixed & Controlled assisted-path validation on contextual attacks and borderline benign quotations \\
Benign chat & 3 & Benign & Low-friction pass-through on ordinary user requests \\
Benign web content & 2 & Benign & Resistance to false positives on retrieved content \\
Benign tool use & 15 & Benign & Policy-allowed proxy forwarding \\
Keyword false-positive controls & 2 & Benign & Oversensitivity stress test for surface-form heuristics \\
\bottomrule
\end{tabularx}
\caption{Current preliminary benchmark corpus. The table is intentionally small and should be read as a pilot benchmark snapshot rather than as the final evaluation scale.}
\label{tab:seed-corpus}
\end{table}

For artifact reproduction, the default workflow requires two commands:

\begin{verbatim}
corepack pnpm -r build
corepack pnpm --filter @kyaclaw/cli exec tsx ../../evaluation/run-benchmark.ts
\end{verbatim}

The resulting report is stored in \texttt{evaluation/results/benchmark-report.json}, with a human-readable summary in \texttt{evaluation/results/seed-run-summary.md}.

To keep scanner-path evidence reproducible, the harness also records run-level scanner metadata, including whether a run was executed in default, controlled mock-assisted, or explicitly live-model mode, together with the effective model label and timeout budget. This metadata matters because a live-model experiment should not silently degrade into heuristic fallback while still being described as an assisted-path run.

\subsection{Attack Suites and Benign Controls}

Our evaluation is organized by attack surface rather than by a single prompt-injection label. This design choice reflects the fact that PRISM is a runtime defense architecture whose mechanisms are distributed across prompt processing, tool invocation, outbound messaging, and policy enforcement---collapsing all attacks into a single category would obscure which layer is responsible for which outcome.

The current harness instantiates the following suite families:

\begin{itemize}[leftmargin=1.5em]
  \item \textbf{Direct injection}: attempts to override instructions, take over agent roles, force system-prompt disclosure, or exploit format-token boundaries.
  \item \textbf{Indirect injection}: malicious text plausibly delivered through fetched web content, including percent-encoded directives and zero-width-character obfuscation.
  \item \textbf{Exfiltration}: attempts to upload, disclose, or transmit tokens, secrets, or API keys, including command-oriented patterns such as piping to a remote shell.
  \item \textbf{Tool abuse}: unauthorized callers, foreign-session access attempts, invocations of disallowed tools, default-deny tool requests, and dangerous shell-trampoline commands.
  \item \textbf{Benign chat and web}: natural-language requests and web-retrieval results that should pass through without triggering any defense.
  \item \textbf{Benign tool use}: proxy-policy requests that should be allowed under the configured policy.
  \item \textbf{Keyword false-positive controls}: benign text deliberately containing security-related vocabulary intended to measure oversensitivity in surface-form heuristics.
\end{itemize}

In the full evaluation, these suites will be expanded with additional file-tampering cases, a larger indirect-injection corpus, and end-to-end hook-level scenarios. The guiding design principle is to preserve suite identity so that blocking rates and false positives can be attributed to specific defense mechanisms rather than collapsed into a single aggregate number.

\subsection{Baselines and Ablations}

Because PRISM is explicitly a layered system, a single detector-style baseline is insufficient to demonstrate why lifecycle-wide interception, proxy enforcement, or session-level risk accumulation matter. The evaluation therefore compares multiple deployment configurations arranged as an incremental baseline ladder:

\begin{enumerate}[leftmargin=1.5em]
  \item \textbf{No PRISM} --- the unprotected OpenClaw gateway.
  \item \textbf{Heuristics only} --- local text heuristics with no sidecar services.
  \item \textbf{Plugin only} --- all ten lifecycle hooks active, but no external scanner, proxy, or monitor.
  \item \textbf{Plugin + scanner} --- lifecycle hooks plus the scanner cascade.
  \item \textbf{Full PRISM} --- all components active, including proxy, monitor, dashboard, and DLP.
\end{enumerate}

This ladder allows each configuration to be compared against both the unprotected baseline and the immediately preceding level, isolating the incremental contribution of each layer.

For ablation experiments, the most important single-component removals are:

\begin{itemize}[leftmargin=1.5em]
  \item Removing the remote scan cascade (heuristic-only classification).
  \item Removing session risk accumulation (each message judged independently).
  \item Removing the proxy and exec-governance layer (tool calls forwarded without policy checks).
  \item Removing outbound DLP (credential patterns not scanned in outbound messages).
  \item Removing the file monitor (control-file changes not detected).
\end{itemize}

The current preliminary harness directly supports only a subset of this design space. Specifically, it exercises heuristics-only behavior through the heuristic engine, scanner-path behavior through the scanner engine, policy-layer behavior through the proxy-policy engine, and hook-level lifecycle behavior through the plugin-only, plugin-scanner, and full-PRISM engines. These partial engines are useful for early validation but are not substitutes for the full end-to-end baseline comparison.

\subsection{Metrics}

We report two complementary families of metrics: security metrics and systems metrics.

\paragraph{Security metrics.} For suites dominated by text classification and policy enforcement, the harness computes accuracy, precision, recall, F1, attack block rate, and false positive rate. These metrics are already emitted in the JSON report and are sufficient for the preliminary runs presented in Section~5.5.

\paragraph{Systems metrics.} Because PRISM is a runtime defense layer rather than an offline classifier, security efficacy alone is an incomplete picture. The full evaluation must additionally report per-hook latency (p50, p95, p99), end-to-end latency delta between protected and unprotected configurations, CPU and memory overhead disaggregated by component, audit-chain verification success rate and tamper-detection rate, configuration-reload latency, and allow-action completion latency.

The current harness now emits a preliminary profiling block for each engine. It records p50/p95/p99 case timing, CPU time per executed case, and peak RSS delta. We also supplement the report with narrow component-level operational timings for audit verification and proxy policy reload. These measurements are useful for current benchmark characterization, but they are still weaker than a deployment-level overhead study across real gateway and sidecar processes.

\subsection{Preliminary Benchmark Results}

The results presented in this subsection should be read as preliminary benchmark evidence rather than as final experimental findings. At the current stage we maintain two benchmark families. The first is a text-and-policy corpus with default and controlled mock-assisted scanner runs. The second is a step-sequenced \texttt{plugin-flow} corpus intended to ground the middle rows of the baseline ladder before a full OpenClaw gateway harness exists.

In the \textbf{default benchmark run}, the \textbf{no-PRISM engine} executes all 110 preliminary cases and correctly classifies only the 50 benign ones, yielding accuracy 0.455 and attack recall 0.000. This deliberately weak lower bound is useful because it anchors the baseline ladder to a concrete ``forward-everything'' reference rather than leaving the unprotected row purely conceptual. On the protected text path, the \textbf{heuristic engine} and the \textbf{scanner engine} each execute 30 \texttt{scan-text} cases and both achieve 15 out of 30 correct classifications, yielding accuracy 0.500, precision 0.526, recall 0.625, and F1 0.571. The \textbf{proxy-policy engine} now executes 33 \texttt{invoke-policy} cases and classifies all 33 correctly. Scanner-path telemetry shows 8 \texttt{heuristic-shortcircuit} cases and 22 \texttt{heuristic-fallback} cases, with zero \texttt{ollama-assisted} cases. This default run is therefore useful primarily as a transparency check: absent an available model endpoint, the scanner engine degenerates to heuristic behavior rather than implicitly claiming additional coverage.

In the \textbf{controlled mock-assisted run}, the harness enables \texttt{--scanner-ollama-mock}, which serves case-embedded model verdicts only for explicitly annotated \texttt{scanner-focused} cases. Under this configuration, the \textbf{no-PRISM engine} remains unchanged at 50 out of 110 because it never consults the scanner, the heuristic engine remains at 15 out of 30 correct classifications, but the scanner engine improves to 26 out of 30, yielding accuracy 0.867, precision 0.800, recall 1.000, and F1 0.889. The proxy-policy engine remains at 33 out of 33 because it does not depend on the scanner path. Scanner-path telemetry records 13 \texttt{ollama-assisted} cases, 8 \texttt{heuristic-shortcircuit} cases, and 9 \texttt{heuristic-fallback} cases. The main value of this run is not to claim real model performance, but to demonstrate that the benchmark can isolate and measure scanner-path differences when those paths are explicitly exercised.

The structure of the remaining errors is informative. In the mock-assisted run, the scanner resolves all of the newly added contextual attack cases and all of the new quoted benign controls in the \texttt{scanner-focused} suite, but it still misclassifies four high-score benign controls: the two \texttt{keyword-false-positive-controls} cases and the two earlier \texttt{scanner-focused} training/reporting cases whose heuristic scores exceed the short-circuit threshold. This residual error pattern exposes an important limitation of the current cascade design: once the local heuristic score crosses the high-risk short-circuit boundary, the model path is never consulted.

These runs should therefore be interpreted conservatively. The default run validates transparent fallback behavior, and the controlled mock-assisted run validates path instrumentation and table generation under controlled conditions. Neither run should be presented as a substitute for a live-model evaluation under fixed model, timeout, and hardware settings.

The expanded \textbf{plugin-flow} benchmark family now provides 47 lifecycle cases inside the unified baseline-ladder slice. On this suite alone, \textbf{plugin-only} reaches 34 out of 47 correct outcomes in the default, mock-assisted, and live runs. \textbf{Plugin-scanner} improves to 38 out of 47 in the default run, 41 out of 47 in the controlled mock-assisted run, and 40 out of 47 in the live local-model run. \textbf{Full-PRISM} matches the default and mock lifecycle totals at 38/47 and 41/47 and, in the current live benchmark, matches \textbf{plugin-scanner} at 40/47 before adding policy-layer coverage on the \texttt{tool-abuse} and \texttt{tool-use} suites. The most informative differential cases remain contextual tool-result attacks that only become blockable after scanner-assisted risk escalation, together with quoted benign training examples that plugin-only still over-blocks because scanner failure contributes a fallback risk bump.

To reduce the remaining comparability gap between rows in the baseline ladder, we also maintain a \textbf{unified baseline-ladder slice} consisting of the \texttt{plugin-flow}, \texttt{tool-abuse}, and \texttt{tool-use} suites. This slice now contains 80 cases and is the strongest current same-slice benchmark in which all five target configurations---\textbf{No PRISM}, \textbf{Heuristics only}, \textbf{Plugin only}, \textbf{Plugin + scanner}, and \textbf{Full PRISM}---are computed over the same case set. In the \textbf{default unified run}, these five rows achieve 36/80, 46/80, 49/80, 53/80, and 71/80 correct outcomes, respectively. In the \textbf{controlled mock-assisted unified run}, the first three rows remain unchanged while \textbf{Plugin + scanner} improves to 56/80 and \textbf{Full PRISM} to 74/80. We then repeated the same unified slice in a \textbf{live local-model run} using a local Ollama endpoint with \texttt{qwen3:8b}, a 20\,s model timeout, fixed-temperature local decoding, and a plugin-side scanner timeout widened to 21\,s to avoid mistaking local-model latency for scanner failure. Under this live setting, the harness records more than a dozen genuinely model-assisted cases; \textbf{Plugin + scanner} reaches 55/80 and \textbf{Full PRISM} reaches 73/80, while \textbf{No PRISM}, \textbf{Heuristics only}, and \textbf{Plugin only} remain at 36/80, 46/80, and 49/80. The corresponding live attack block rates are 0.000, 0.409, 0.455, 0.545, and 0.955, while the false-positive rates are 0.000, 0.222, 0.194, 0.139, and 0.139. This live benchmark remains preliminary and should not be confused with the final end-to-end table, but it is a materially stronger paper-facing reference point than the earlier mock-assisted slice because it confirms that the scanner lift is observable under a real local model rather than only under benchmark-injected verdicts.

The same live benchmark now also carries a \textbf{preliminary harness-level overhead profile}. On the current 80-case live slice, the unprotected and local-only configurations remain sub-millisecond at p95 (0.007\,ms for \textbf{No PRISM}, 0.046\,ms for \textbf{Heuristics only}, and 0.583\,ms for \textbf{Plugin only}), whereas the scanner-backed rows rise sharply to 12.5\,s for \textbf{Plugin + scanner} and 15.8\,s for \textbf{Full PRISM}. This gap is unsurprising: on the preliminary slice, the dominant cost is local-model scanner latency on the subset of cases that exercise the remote classification path, not local heuristic processing. Peak RSS deltas in this single-process harness stay below 1.4\,MiB across the five rows, indicating that the visible cost at this stage is primarily latency rather than memory growth. We also measured two component-level operational timings: on a 200-entry synthetic audit log, p95 audit-chain verification is 4.683\,ms for chain-only verification and 8.846\,ms when anchor verification is included, while proxy \texttt{reloadPolicy()} reaches p95 0.574\,ms. These figures are useful for preliminary benchmark characterization and for grounding the discussion of recoverability, but they should not be interpreted as substitutes for a deployment-level end-to-end overhead study.

\begin{table}[t]
\centering
\footnotesize
\begin{tabularx}{\textwidth}{l c c c c c}
\toprule
Configuration & Attack block rate & False positive rate & Precision & Recall & F1 \\
\midrule
No PRISM & 0.000 & 0.000 & n/a & 0.000 & n/a \\
Heuristics only & 0.409 & 0.222 & 0.692 & 0.409 & 0.514 \\
Plugin only & 0.455 & 0.194 & 0.741 & 0.455 & 0.563 \\
Plugin + scanner & 0.545 & 0.139 & 0.828 & 0.545 & 0.658 \\
Full PRISM & 0.955 & 0.139 & 0.894 & 0.955 & 0.923 \\
\bottomrule
\end{tabularx}
\caption{Preliminary unified baseline-ladder results on the current 80-case same-slice benchmark consisting of \texttt{plugin-flow}, \texttt{tool-abuse}, and \texttt{tool-use} suites under a live local-model scanner setting (\texttt{qwen3:8b} via Ollama). This is the strongest current five-row comparison, but it remains a preliminary benchmark rather than the final end-to-end paper result.}
\label{tab:prelim-baseline-ladder}
\end{table}

\begin{table}[t]
\centering
\footnotesize
\begin{tabularx}{\textwidth}{l c c c c c}
\toprule
Configuration & p50 ms & p95 ms & p99 ms & CPU ms / case & Peak RSS $\Delta$ MiB \\
\midrule
No PRISM & 0.002 & 0.007 & 0.209 & 0.000 & 0.027 \\
Heuristics only & 0.005 & 0.046 & 2.297 & 0.000 & 0.219 \\
Plugin only & 0.083 & 0.583 & 3.619 & 0.200 & 1.348 \\
Plugin + scanner & 0.178 & 12498.437 & 19874.211 & 10.162 & 0.066 \\
Full PRISM & 1.799 & 15774.620 & 20045.816 & 21.087 & 0.980 \\
\bottomrule
\end{tabularx}
\caption{Preliminary harness-level overhead profile on the live 80-case baseline-ladder artifact. These numbers reflect benchmark-path timing and local-process resource cost, not deployment-level service latency.}
\label{tab:prelim-overhead}
\end{table}

\begin{table}[t]
\centering
\footnotesize
\begin{tabularx}{\textwidth}{l c c c c c c}
\toprule
Operation & Avg ms & p50 ms & p95 ms & p99 ms & Min ms & Max ms \\
\midrule
Verify chain only & 2.932 & 2.483 & 4.683 & 4.683 & 1.804 & 4.683 \\
Verify chain + anchors & 5.308 & 4.882 & 8.846 & 8.846 & 3.856 & 8.846 \\
Proxy reloadPolicy() & 0.345 & 0.336 & 0.574 & 0.574 & 0.241 & 0.574 \\
\bottomrule
\end{tabularx}
\caption{Preliminary component-level operational timings used to ground PRISM's recoverability claims. The audit timings are measured on a synthetic 200-entry log with anchors every 10 entries; the reload timing measures proxy policy reload on a temporary local policy file.}
\label{tab:prelim-operational}
\end{table}

\subsection{Case Studies}

In addition to aggregate metrics, the evaluation will include short case studies that illustrate how specific PRISM mechanisms respond to representative attack vectors. These studies serve a different purpose than the quantitative tables: they demonstrate the system's behavior at the level of individual interactions rather than statistical summaries.

The target case studies are:

\begin{itemize}[leftmargin=1.5em]
  \item \textbf{Case 1}: indirect injection embedded in fetched web content; detected by post-tool scanning and redacted at persistence/write stages; residual limitation depends on heuristic coverage of novel injection patterns.
  \item \textbf{Case 2}: shell-command abuse via trampoline (\texttt{curl | sh}, \texttt{bash -c}); blocked before execution by command-parsing rules; residual limitation is that novel trampoline patterns may require rule updates.
  \item \textbf{Case 3}: outbound credential leakage in an agent response; blocked by DLP pattern matching at the message-sending stage; residual limitation is coverage of only known credential prefixes.
  \item \textbf{Case 4}: gradual escalation through repeated low-grade signals; risk score accumulates across turns and high-risk tools are blocked at threshold; residual limitation depends on TTL window and attacker pacing.
  \item \textbf{Case 5}: cross-session contamination via a shared channel; no contamination because risk state is split between conversation-scoped and session-scoped keys rather than shared channel IDs; residual limitation depends on correct session-key propagation by the gateway.
  \item \textbf{Case 6}: audit log tampered after the fact; chain break detected by HMAC and hash verification; residual limitation is that detection is post-hoc rather than preventive.
\end{itemize}

This format ensures that each case study includes not only the defense outcome but also the residual limitation, preventing the section from drifting into uncritical success narratives.

\section{Related Work}

Work related to PRISM falls into four broad categories: prompt-injection and jailbreak defense, security mechanisms for tool-using agents, runtime monitoring and policy enforcement, and tamper-evident audit mechanisms. PRISM is closest to work that treats agent security as a runtime systems problem rather than as a standalone text-classification task. At the same time, PRISM does not claim to be the first secure-agent framework or the first guardrail system. Its distinction is narrower and more concrete: a zero-fork runtime security layer for OpenClaw-based agent gateways that combines lifecycle-wide interception, staged risk response, policy-enforced tool and network controls, and an operational audit plane within a single deployable system.

\subsection{Prompt-Injection and Boundary Guardrails}

One important line of work studies prompt injection and jailbreak defense primarily at the level of text classification or boundary filtering. Early work by Greshake et al.~\cite{greshake2023} established indirect prompt injection as a concrete threat to LLM-integrated applications by showing how adversaries can plant malicious instructions in retrieved content rather than in direct user prompts. Survey-style work such as \emph{Security of AI Agents}~\cite{securityofaiagents2024} provides a broader view of the attack surface introduced by agentic systems---including prompt override, unsafe tool use, and data leakage---and helps motivate why prompt-only defenses are insufficient in agent settings. Benchmark-oriented work such as \emph{Agent Security Bench (ASB)}~\cite{asb2024} and large-scale prompt-hacking studies such as \emph{HackAPrompt}~\cite{hackaprompt2023} further highlight that attacks can manifest across multiple stages of agent operation rather than at a single input boundary.

More targeted defenses have explored firewall-style protection against indirect prompt injection. \emph{Indirect Prompt Injections: Are Firewalls All You Need, or Stronger Benchmarks?}~\cite{indirectpromptfirewalls2025}, for example, frames defense in terms of a tool-input firewall (Minimizer) and a tool-output firewall (Sanitizer) placed around untrusted content flows. PRISM shares the same concern that indirect injection often enters through fetched content or tool-returned text, but differs in scope: rather than treating tool input and tool output as the only security boundaries, PRISM distributes enforcement across message ingress, prompt construction, pre-execution tool checks, post-tool scanning, persistence-time redaction, outbound filtering, and session-lifecycle hooks. The relevant comparison is therefore not that PRISM replaces boundary firewalls, but that it generalizes boundary filtering into a broader runtime-defense model for tool-augmented gateways.

\subsection{Security Mechanisms for Tool-Using Agents}

A second line of work focuses on guardrails designed specifically for LLM agents that use tools. \emph{LlamaFirewall}~\cite{llamafirewall2025} presents an open-source guardrail framework for secure AI agents and represents one of the closest points of comparison to PRISM. \emph{NeMo Guardrails}~\cite{nemoguardrails2023} likewise provides programmable runtime rails for controllable and safe LLM applications. Both systems treat agent security as an operational problem rather than a single detection benchmark. The primary difference lies in framing and integration depth: these systems are presented as more general guardrail frameworks, whereas PRISM is deliberately implemented as a zero-fork runtime layer attached to one concrete gateway stack. This narrower scope allows PRISM to lean more heavily on gateway lifecycle hooks, deploy-time policy enforcement, and operator workflows, at the cost of reduced framework generality.

Other work employs dedicated guard agents or adaptive safety mechanisms to supervise agent behavior. \emph{GuardAgent}~\cite{guardagent2024} proposes a guard-agent approach based on knowledge-enabled reasoning, while \emph{AGrail}~\cite{agrail2025} emphasizes adaptive and lifelong safety checks for agent tasks. These approaches are complementary to PRISM but reflect different design philosophies. Their primary emphasis is on supervisory reasoning or adaptive safety detection; PRISM instead emphasizes explicit interception points, policy-backed runtime controls, and recoverable operational behavior. PRISM is not built around the assumption that a single supervisory model can reliably arbitrate all unsafe behavior; it is built around the premise that agent security benefits from multiple concrete enforcement points tied directly to the runtime itself.

Recent work on safe tool use makes a distinct but highly relevant contribution. \emph{Towards Verifiably Safe Tool Use for LLM Agents}~\cite{safe-tool-use2026} argues for deriving enforceable safety requirements from structured workflow analysis and formalized tool constraints. This direction is more ambitious in terms of formal assurance than PRISM currently pursues. PRISM should therefore be understood as complementary rather than competing on the same axis: it provides deployable runtime controls and auditing for an existing gateway implementation, while formal safe-tool-use methods point toward a stronger future direction for specifying and verifying tool-level safety properties.

\subsection{Runtime Monitoring, Policy Enforcement, and Benchmarks}

A third category encompasses work that treats agent security as a runtime-control or evaluation problem. Some of this literature focuses on formalizing attacks and defenses, while other work focuses on adaptive guardrails or workflow-level constraints. \emph{Agent Security Bench (ASB)}~\cite{asb2024} is particularly relevant here because it frames agent security across multiple scenarios, tools, and attack types, demonstrating that vulnerabilities span prompt handling, tool usage, and memory-related stages. \emph{ToolEmu}~\cite{toolemu2023} is also relevant because it provides a way to evaluate tool-using agents against high-stakes scenarios without fully instantiating each external tool environment. PRISM is not a competing benchmark; rather, it is a deployable defense architecture that could in principle be evaluated using benchmark methodologies of this kind.

The most important conceptual distinction in this category is between detector-style guardrails and policy-enforced runtime defense. Detector-style systems primarily answer whether a prompt or tool result appears suspicious. PRISM includes detection as well, but it couples detection with specific enforcement controls: high-risk tool blocking, protected-path enforcement, private-network restrictions, tiered-domain handling, outbound secret filtering, and operator-mediated exception workflows. This coupling is significant because many agent failures are not best characterized as classification errors alone---they are failures of execution policy, state management, or unsafe privilege exposure. In that sense, PRISM is closer to a runtime security layer than to a pure prompt-defense classifier.

\subsection{Tamper-Evident Auditing and Operational Security}

A fourth related line of work originates from the secure-logging and tamper-evident auditing literature. Prior systems outside the agent-security domain have long studied append-only logs, chained integrity records, and post-hoc verification mechanisms for detecting unauthorized modification of historical events; Schneier and Kelsey~\cite{schneierkelsey1998}, for example, provide an early cryptographic formulation of secure logging on untrusted machines. PRISM does not claim a new cryptographic audit primitive. Instead, it adapts established tamper-evident techniques to an agent-runtime setting by combining HMAC-protected audit entries, chained hashes, anchor-based verification, and CLI verification tooling with a runtime dashboard and hot-reloadable policy plane.

This operational coupling is an important distinction from detector-only guardrail systems. In PRISM, auditability is not a secondary observability feature added after detection; it is integral to the security argument for recoverability, operator review, and policy-change management. The same holds for health probes, configuration reloads, and allow workflows. These features are not algorithmic contributions in the model-research sense, but they help distinguish PRISM from work that primarily studies detection quality while leaving runtime operations outside the system boundary.

Overall, PRISM sits at the intersection of these four threads: it is broader than boundary-only prompt firewalls, more operationally grounded than detector-only guard agents, less formal than specification-driven safe-tool-use work, and more agent-specific than generic tamper-evident logging systems. The resulting contribution is not a new classifier or a new cryptographic primitive, but a deployable runtime security architecture that integrates multiple defense and operations mechanisms into a single gateway-facing system.

\section{Discussion and Limitations}

PRISM is designed as a deployable runtime security layer, not as a complete security boundary for LLM agents. Its primary contribution is to distribute enforcement across multiple runtime stages and to connect detection, policy enforcement, auditing, and operator workflows within a zero-fork gateway integration. That design provides practical value for agent gateways, but it also introduces a clear set of boundaries that should be stated explicitly.

\paragraph{Detection coverage is necessarily incomplete.} The heuristic layer is intentionally fast and operationally simple, which makes it suitable for inline use across multiple hooks, but it cannot cover all prompt-injection variants, obfuscation strategies, or context-dependent malicious instructions. The optional LLM-assisted scanner broadens coverage for suspicious tool-returned content, yet it remains a best-effort classifier rather than a formally trustworthy arbiter. Adaptive or novel adversarial inputs may evade both tiers. PRISM should therefore be understood as a layered runtime defense that raises the difficulty of successful attacks and narrows the exposure window, not as a mechanism that can guarantee perfect detection of all adversarial inputs.

\paragraph{Safety properties depend in part on explicit runtime policy, not only on classification.} This is a strength, because many failures of tool-using agents are better described as policy failures than as text-understanding failures. It is also a limitation, because policy quality must be maintained over time. Allowlists, denylists, protected-path rules, domain tiers, and DLP patterns all require review as deployment needs evolve. PRISM therefore reduces operational security burden in some areas by centralizing controls, but it does not eliminate the need for ongoing policy maintenance. Operators who deploy PRISM still bear responsibility for the correctness and currency of their configured policy surface.

\paragraph{File and path protections are intentionally scoped.} The current path-canonicalization logic performs string-level normalization and resolution, but it does not resolve symlink aliases through \texttt{realpath}. As a result, the system can enforce practical path checks for the common cases it models directly, but it should not be described as a complete filesystem sandbox. Likewise, the file-integrity monitor and audit chain are designed to make tampering visible and verifiable after the fact; they provide tamper-evident behavior, not tamper-proof storage. An attacker who can bypass string-level path resolution or who operates below the filesystem abstraction layer falls outside PRISM's protection boundary.

\paragraph{PRISM is framework-specific by design.} The zero-fork integration strategy depends on the hook surface and deployment assumptions of OpenClaw-based gateways. This specificity is part of why the implementation can be concrete and deployable, but it also limits direct portability. Applying the same architectural approach to other agent frameworks would likely preserve the overall defense model while requiring substantial re-mapping of lifecycle stages, tool abstractions, policy insertion points, and operational interfaces. The degree to which PRISM's mechanisms generalize beyond OpenClaw remains an open question that can only be answered through future porting efforts.

\paragraph{PRISM does not replace lower-layer system hardening.} It does not provide OS-level sandboxing, kernel isolation, hardware-rooted integrity, or general outbound network control outside the paths it directly mediates. A secure deployment should therefore treat PRISM as one layer in a broader defense stack that may also include host hardening, container or VM isolation, network egress controls, secret management infrastructure, and conventional service authentication. The absence of any of these complementary layers may leave attack surfaces that PRISM, by design, does not address.

\paragraph{The current system should not be overinterpreted as production-proven at large scale.} The implementation is real, runnable, and test-covered, and the evaluation methodology is designed to measure effectiveness, false positives, layer contribution, overhead, and operational recovery. Even so, large-scale multi-tenant validation, long-duration operational studies, and broader benchmark coverage remain future work. Claims about deployability are therefore strongest at the level of architecture and implementation maturity, not at the level of broad field validation.

\subsection{Future Directions}

These limitations also point toward several natural extensions. A productive next step would be to expand benchmark coverage with larger indirect-injection and tool-abuse corpora, then evaluate the system under more diverse workloads and deployment settings. A second direction is to combine PRISM's runtime hooks and policy layer with more formal safe-tool-use specifications, allowing deployable enforcement to be paired with stronger pre-deployment reasoning about tool constraints. Porting the architecture to other agent frameworks would help separate what is intrinsic to the PRISM design from what is specific to OpenClaw's lifecycle model. Finally, audit-driven policy refinement may provide a practical path toward improving rules and thresholds based on observed attack and false-positive patterns over time, closing the loop between runtime telemetry and policy evolution.

Overall, the appropriate interpretation of PRISM is not that it solves agent security in full, but that it demonstrates a practical systems approach for advancing agent defenses from isolated filters toward a lifecycle-aware, policy-enforced, and auditable runtime layer.

\section{Conclusion}

Tool-augmented LLM agents expose a security surface that extends well beyond user-input filtering. In practical gateway deployments, risk emerges across message ingress, prompt construction, tool invocation, tool-returned content, outbound messaging, and local control files. These distributed attack surfaces render single-boundary defenses insufficient for many real agent workflows, particularly when prompt injection, tool abuse, credential leakage, and stateful escalation can interact across multiple runtime stages.

This paper has presented OpenClaw PRISM, a zero-fork runtime security layer for OpenClaw-based agent gateways. Rather than proposing a new detection model, PRISM combines ten lifecycle hooks, a hybrid heuristic-plus-LLM scanning pipeline, conversation-scoped and session-scoped risk accumulation, policy-enforced controls over tools, paths, and outbound traffic, and a tamper-evident audit and operations plane. The resulting architecture aims to advance agent security from isolated input filtering toward a deployable runtime layer that can be attached to an existing gateway, supervised as a set of long-running services, audited after the fact, and adjusted through explicit operational workflows---all without forking the upstream framework.

The broader message of this work is that agent security should be treated as a systems problem as much as a detection problem. Runtime interception, staged policy response, audit integrity, and recoverable operator workflows all matter when security controls are expected to function inside long-running, tool-using gateways. PRISM is one concrete instantiation of that perspective: not a complete defense, not a formal guarantee, and not a substitute for lower-layer hardening such as OS sandboxing or network-level egress control, but a practical architecture for making agent gateways more controllable, observable, and defensible in deployment.

Several directions remain open for future work. Stronger benchmark coverage---especially for indirect injection, tool abuse, and long-horizon escalation---would improve the evidence base for runtime defenses of this kind. Extending the architecture beyond the OpenClaw ecosystem would help clarify which mechanisms are framework-specific and which generalize to other agent runtimes. A particularly promising direction is to pair PRISM's deployable hook and policy infrastructure with more formal safe-tool-use specifications or with stronger audit-driven policy refinement. Together, these directions point toward a broader agenda in which agent security is constructed from layered runtime controls, explicit policy boundaries, and operationally verifiable system behavior.

\bibliographystyle{plain}
\bibliography{references}

\end{document}